\documentclass[pre,showpacs,amsmath,amssymb,reprint,superscriptaddress, nofootinbib]
{revtex4-1}

\usepackage{graphicx}
\usepackage{amsmath}
\usepackage{array}
\usepackage{color}
\usepackage{hyperref}

\newcommand{\eg}{{e.g., }}
\newcommand{\etal}{{\it et al. }}
\newcommand{\ie}{{\it i.e., }}

\newcolumntype{M}{>{\centering\arraybackslash}m{\dimexpr.28 \linewidth-1\tabcolsep}}

\begin{document}

\title{A shapeable material without plastic deformation} 
\author{Naomi Oppenheimer} 
\email{naomiop@gmail.com}
\affiliation{James Franck Institute, University of
  Chicago, Chicago, IL 60637, USA}

\author{Thomas A. Witten}
\email{t-witten@uchicago.edu} 
\affiliation{James Franck Institute, University of
  Chicago, Chicago, IL 60637, USA}

\date{\today}

\begin{abstract}

Randomly crumpled sheets have shape memory. In order to understand the basis of this form of memory, we simulate triangular lattices of springs whose lengths are altered to create a topography with multiple potential energy minima.  We then deform these lattices into different shapes and investigate their ability to retain the imposed shape when the energy is relaxed. The lattices are able to retain a range of curvatures. Under moderate forcing from a state of local equilibrium, the lattices deform by several percent but return to their retained shape when the forces are removed. By increasing the forcing until an irreversible motion occurs, we find that the transitions between remembered shapes show co-operativity among several springs. For fixed lattice structures,  the shape memory tends to decrease as the lattice is enlarged; we propose ways to counter this decrease by modifying the lattice geometry. We survey the energy landscape by displacing individual nodes. An extensive fraction of these nodes proves to be bistable; they retain their displaced position when the energy is relaxed.  Bending the lattice to a stable curved state alters the pattern of bistable nodes.  We discuss this shapeability in the context of other forms of material memory and contrast it with the shapeability of plastic deformation.  We outline the prospects for making real materials based on these principles.

\end{abstract}


\maketitle

\section{introduction}
\label{sec_intro}

If we take a piece of paper and pressure it from both sides, it will form one large buckle. Once the pressure is released, the paper will go back to being flat. We now repeat the experiment, but first crumple the piece of paper to a little ball, open, and flatten it then apply pressure again. When we now release the pressure, the paper will retain some curvature. Not only that, but it can be shaped in various forms which are somewhat stable to an applied force. 

One evident source of this shapeability is the local plasticity\cite{Bedia2011} of paper.  Each fold produced by the crumpling process has undergone a permanent structural change in the paper's fiber matrix. Moreover, the resulting ridges and vertices store memory and create an intricate landscape that has many metastable configurations \cite{Lobkovsky1995}. Thus reshaping it into a different crumpled form causes a crackling sound \cite{Kramer1996}, \cite{Houle1996}, as the sheet snaps from one metastable minimum to another. 
 
In this work we generalize this effect to an {\it elastic} sheet. The purpose is twofold, first to create a material that is reshapeable and stable. Second to understand the origin of shapeability in a simple realization, one in which there are only Hookean springs and thus no plastic deformation.  
Our approach is to use an array of springs of varying rest lengths in a geometry that creates many locally stable, interacting configurations.
We use two different models --- the {\it random lattice}, having randomness in the springs' rest length, and the {\it puckered lattice} which is a regular structure, and has the same repeating unit throughout the lattice. We study properties of the zero-temperature ground states of our lattice. This is separate from the well-studied co-operativity of thermally fluctuating ``tethered" lattices of Kantor, Kardar and Nelson~\cite{Kantor1986},\cite{Kantor1987}.

Before we go on to explain each model, we define the quality of shapeability that we intend to explore. A shapeable material is one that deforms under external forces, and which retains its deformed shape when these forces are removed.  Moreover this retained shape is stable:  when further deformed by sufficiently mild forces, the object returns to the retained shape.

The basic feature that enables shapeability in a crumpled sheet is metastability --- the object has many discrete, stable configurations, separated by energy barriers. Metastability in materials and its connection to memory storage is a well-explored field \cite{Mullin2007}--\cite{Cohen2014}. 
Here we survey various forms of shape memory, to distinguish these from the shape memory of a crumpled sheet. One form of shape memory is that of a plastically deforming material such as modeling clay.  As noted above, a simple fold in a sheet of paper is an example of plastic memory.  Setting a shape requires irreversible changes in the microscopic structure within the material. Our aim is to identify a further form of shapeability in {\it crumpled} paper that goes beyond this simple plasticity. 

A second type of shape memory is seen in elastic systems that can switch between two possible states. A simple example is found in a children's toy called the ``slap bracelet", a straight metal strip that wraps itself around a wrist when bending is initiated \cite{SlapBraceletCatalogue}. This piece of metal has positive curvature along one direction and negative along the perpendicular direction. If the strip is sufficiently thin, it will have two cylindrical configurations. It is possible to snap from one to the other using external force \cite{Chen2012}. Another example is seen in shape-memory alloys \cite{Purnawali2010}. These are pseudoelastic materials, able to deform elastically in response to an external stress, and yet return to their initial shape after heating. 

A third category is seen in materials with elastic deformations that result in not two but many configurations. An example is the flexible drinking straw \cite{FlexibleStraw}, a plastic tube with a corrugated region.  When the straw is bent, these corrugations collapse so that the bend is retained. The total curvature is thus determined by the metastability of the corrugations. The difference between a system like the flexible straw and the shapeable sheet is that the former does not require any cooperativity between bistable points. Instead, the global shape is a simple superposition of the effects of each corrugation.

The models that we treat below appear distinct from the categories sketched above. On the one hand, they do not require plasticity (like a simple fold in paper). On the other hand, the global shape is not a simple superposition of the shape of the building blocks (such as in the flexible straw). Instead, many bistable points seem to work in concert to make a changeable shape. The system described by Waitukaitis and von Hecke \etal \cite{Hecke2014} is similar to ours in those aspects; the difference is that the shapeable sheet doesn't require pre-programing an array of possible shapes.

Metastability has been studied in the context of designing and controlling the properties of metamaterials. Silverberg and Cohen \etal \cite{Cohen2014} showed that flipping bistable corners in a Miura Ori sheet can change the bulk properties of the material. Waitukaitis and von Hecke \etal \cite{Hecke2014} studied the energy landscape of 4-degree vertices and discovered they have a surprisingly large number of stable configurations (up to six), and that tiling a space with them preserves the metastability. Periodic elastomeric structures could also be tuned to control many material properties (see for example \cite{Mullin2007}, \cite{Bertoldi2014}, \cite{Hecke2014b}) such as auxeticity, and elastic and acoustic band propagation.

In what follows we will introduce the two models to be studied, examine what conditions are required for them to be shapeable, explore the possible shapes and investigate where the memory resides in the structure. 

\section{Model}
\label{sec_Model}

The system is a network of nodes connected by springs in the topology of a regular triangular two-dimensional lattice. However, we use springs of different rest lengths so that planar configurations are unstable. We embed this system in three-dimensional space and then seek the positions of the nodes that minimize the spring energy.
If the system is well constrained (\ie has bending energy or extra springs, see discussion in the next section) and the springs are all of equal rest length, there is just one stable configuration --- the nodes lie in a plane. Once one introduces a variety of lengths, it is possible to get more than one minimum. The models explored below are not unique; many variations are possible. The random lattice was chosen because it resembles a crumpled sheet; the puckered lattice was chosen because it is simpler to understand; it also demonstrates shapeability in an ordered material. 

Before going into detail about each model let us describe properties which apply to both of them and to every triangular lattice of springs. Specifically, let us determine when the springs provide enough constraints to dictate specific configurations of nodes. When they do not, what characterizes the modes of deformation that cost no energy, known as the floppy modes? 

\subsection{Floppy modes}
\label{sec_floppy}
Our interest is in systems that hold their shape, \ie rigid objects. In this section we explain why our lattice requires modification in order to hold its shape. Neglecting edge effects, a lattice of $N$ nodes contains $3N$ springs.  Each spring imposes a scalar constraint on the $3N$ node co-ordinates.  Thus the springs are just sufficient to constrain the node positions\cite{Maxwell1864, MaxwellNote}.

However, in a finite lattice of $N$ nodes cut from an infinite lattice, there are fewer than $3N$ springs; the springs that connected the lattice to the infinite lattice have been removed.  This number is proportional to the perimeter.  Thus any finite lattice has a number of unconstrained internal motions that increases with its size.

\begin{figure}[tbh]
\vspace{0.9cm} 
\includegraphics[height=1.5in]{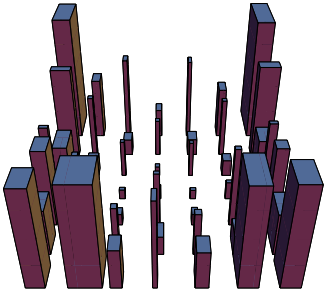} 
\caption{Visualization of the floppy modes in a $7\times7$ puckered lattice of Fig.~\ref{fig_Model}C. Each box shows the three values of floppiness corresponding to the three Cartesian displacements of one node. Thus, the height of the left, rear box (which is $0.98$) shows the floppiness $F_i$ corresponding to the free vertical displacement of that corner node. Likewise, the width and depth of this box are proportional to the floppiness $F_i$ for $x$ and $y$ horizontal displacements. Evidently displacements normal to the lattice have relatively large floppy content.}
\label{fig_Floppy}
\end{figure}

In order to gauge how these floppy modes might compromise shapeability of our lattices, we calculated the modes explicitly. Given an equilibrium state ${\bf r}_{\rm min}$, with energy $E({\bf r}_{\rm min})$ we calculate the dynamical matrix ${\mathbb M}$  given by, ${\mathbb M}_{ij} = \partial_i\partial_j E({\bf{r_{\rm min}}})$, where $\partial_i$ is the derivative with respect to the $i$th of the $3N$ node co-ordinates. This matrix is symmetric and has either zero or positive eigenvalues. Any eigenfunctions $\{{\bf u}\}$ corresponding to the zero eigenvalues other than those corresponding to pure translations and rotations are the floppy modes\cite{MaxwellNote}. These $\{{\bf u_0}\}$ are also the null space vectors of ${\mathbb M}$, \ie the solutions to the equation ${\mathbb M} \cdot {\bf u_0} = {\bf 0}$. Here the $\{{\bf u_0\}}$ are the directions on the energy landscape that have no energetic cost. 
Any set of displacements can be uniquely expressed in terms of a null vector and a non-null vector normal to all the $\{{\bf u_0\}}$'s. This decomposition may be used to measure the contribution of the floppy modes to any given node co-ordinate $u_i$. In particular, the norm of the null part relative to the total norm gives an unambiguous measure of the relative amount of null-space content in that displacement. We denote this quantity, which lies between 0 and 1, by $F_i$.  The resulting amplitudes are plotted in Fig.~\ref{fig_Floppy}. 

As can be seen, the floppiness lives mostly in the perimeter; the middle is hardly affected. One might think that taking larger lattices makes floppiness irrelevant, as it mostly affects the edges. However, increasing the size of the lattice decreases the energy cost of vertical displacements, so that they become indistinguishable from floppy modes. We address this issue in Sec.~\ref{Sec_Shapes}. 

We may eliminate floppy modes and thereby attain the rigid structure we seek by the addition of constraints. We do so either in the form of extra springs at the edges (next-nearest neighbors) or in the form of bending energy, penalizing deviations from flatness, as detailed in Sec.~\ref{Sec_numerics}.  
In the work that follows we will specify which extra constraints are used. It is of course possible to think of other constraints. Those we used have the advantage of being plausible in actual realizations of the sheet. 

\subsection{Random lattice}
\label{sec_random}
Starting from an equilateral triangular lattice (with no extra springs or bending energy), we increment each spring's resting length by a random increment ranging uniformly over ten percent interval.
The equilateral lattice had a zero energy when flat; in the random lattice the flat realization of the system is frustrated and energy is positive. However it can be completely relaxed by letting the springs move into the third dimension. For this moderate randomness, unless violating some geometrical constraint (such as a spring in a triangle being longer than the sum of the other two), it is always possible to relax the energy entirely. 
Like crumpled paper, the relaxed random lattice forms a surface with a highly irregular pattern suggesting shapeability (see Fig.~\ref{fig_Model}A). We investigate this shapeability below. 

\begin{figure*}[tbh]
\vspace{0.9cm} 
\includegraphics[width=5in]{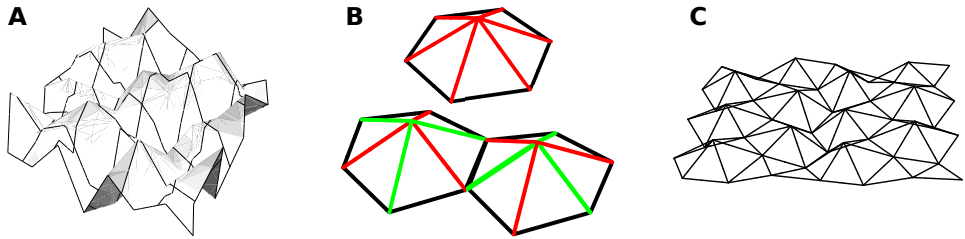} 
\label{fig_Model}
\caption{From left to right: (A) Zero-energy shape of a random lattice of springs with a rest length given by $L=a(1+ 0.1 R[-0.5,0.5])$ where $a$ is the length of a spring in the equilateral case and $R[x,y]$ is a random uniform distribution given values between $x$ and $y$.(B) Top---symmetric case: hexagonal pyramid with equal springs in the middle. Bottom---non-symmetric case: hexagonal pyramid with two different lengths in the middle $a_m=1.05 a$ and $a_l=1.15 a$. Notice that it is no longer symmetric. (C) Zero-energy surface of the puckered lattice with unit cells of  B (bottom).}
\label{fig_Model}
\end{figure*}

\subsection{Puckered lattice}
\label{sec_puckered}
To exhibit shapeability, metastability is required; however the randomness described above is not obligatory. A lattice can have many metastable states with a periodic structure composed of one or more repeating hexagons. One example is a triangular lattice with two different spring rest-lengths as in Fig.~\ref{fig_Model}B(bottom). 
To form it with a simple triangular lattice we lengthen the springs extending from one node to its six neighbors, thus forming a hexagonal pyramid. We then lengthen the six springs at the adjacent hexagons. By extending this process to all the hexagons in the lattice, we may form the lattice of pyramids shown in Fig.~\ref{fig_Model}B(top).

As in the random lattice, a flat configuration is very frustrated. Relieving the frustration results in puckered, hexagonal pyramids. The node in the middle of each pyramid is bistable; it is equally stable above and below the plane of its hexagon. By exerting a sufficient vertical force on such a node, we may ``flip" it through the horizontal plane to the other stable minimum. For a lattice of $N$ nodes there are about $N/3$ bistable nodes, which means $2^{N/3}$ metastable configurations. However, these flips alter the shape only locally. The resulting metastable configurations remain globally flat. The missing ingredient is an energetic coupling between one hexagon and its environment. 
To create the energetic interaction we use springs of three different lengths. The construction is similar to the one just described, except that we add a small mismatch between the lengths of the springs in the middle of each hexagon (see Fig.~\ref{fig_Model}B). In addition, in each column, the spring orientation is rotated by $60^0$, this adds extra frustration. The spring mismatch dictates a shape in which each hexagon is slightly skewed such that it is out of the plane. When flipped, the preferred orientation of the neighbors is modified.   

\section{Shapes}
\label{Sec_Shapes}
Below are a few examples both of equilibrium shapes created with the random lattice and with the puckered lattice.
As seen below, when deformed to match a given ``goal surface", these objects tend to retain the deformed shape when relaxed. That is, the surface defined by the lattice lies close to the  goal surface. In order to fit the lattice optimally to the goal surface we first relax the lattice and determine the area per node $A_r$ of its projection onto its mid-plane  (this projection allows the nodes to be closer to their relaxed density). Then we position the nodes onto a regular triangular lattice with the same area per node in a desired form such as a cylinder. Finally, we move the nodes to find a local energy minimum using standard numerical algorithms, as described in Sec.~\ref{Sec_numerics}. We then compare the resulting shape to the desired form.

\begin{figure*}[t]
\vspace{0.9cm} 
\includegraphics[height=3.5in]{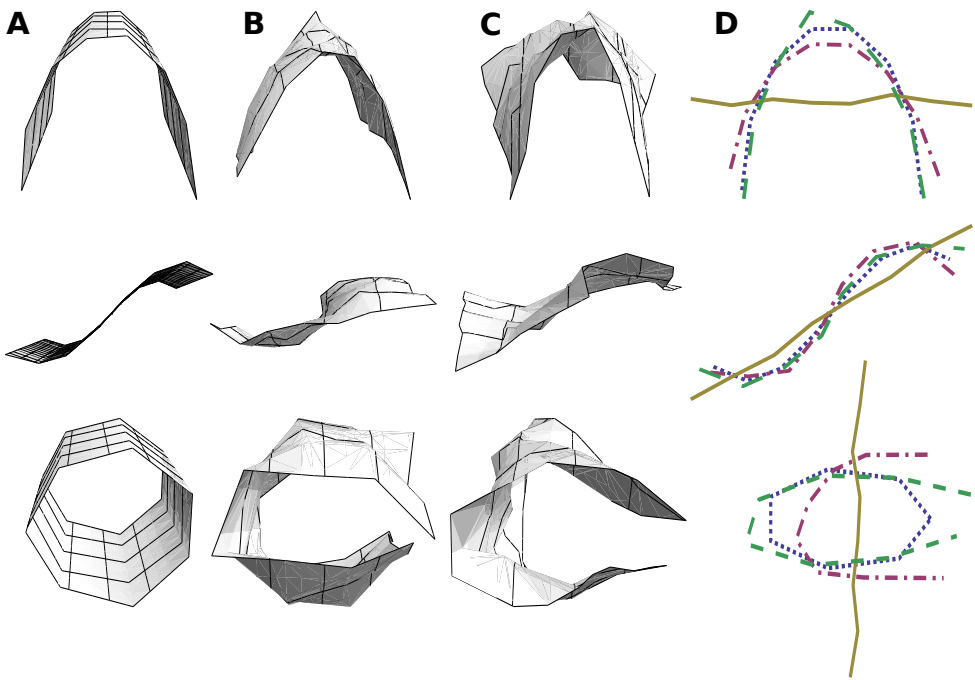} 
\caption{Lattice configurations resulting from the shaping procedure described in the first paragraph of Sec.~\ref{Sec_Shapes}, using a random lattice of size $8 \times 8$. From left to right: (A) Starting from a  goal configuration, all the springs are sitting in a smooth surface but are frustrated; (B) relaxed shapes for a random lattice with bending energy; (C) relaxed shapes for a random lattice with extra springs at the edges; (D) averaged 2D projection of the shapes as described in the text, and a comparison to the original ``flat" sheet: Dotted (blue) is the desired shape, dashed (green) is a sheet with extra springs at the edges, dash-dotted (pink) is a sheet with bending energy, and the yellow is the original ``flat" random sheet. All shapes were translated and rotated so as to get the best fit.} 
\label{fig_shapes_random}
\end{figure*}

\begin{figure*}[tbh]
\vspace{0.9cm} 
\includegraphics[height=1.5in]{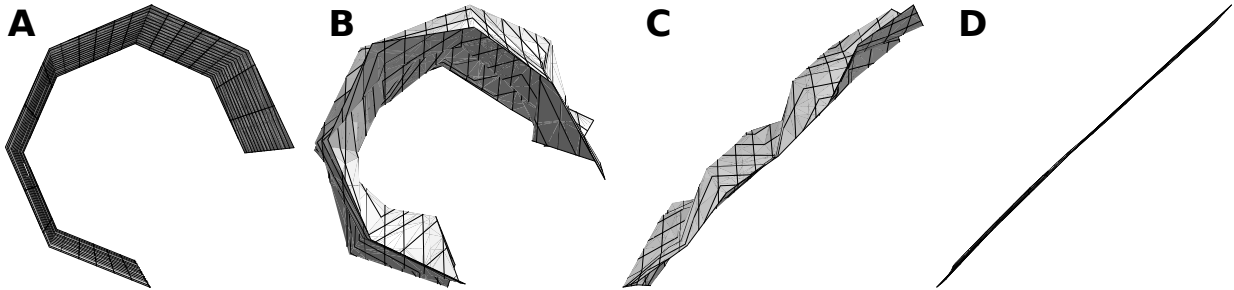} 
\caption{Lattice configurations resulting from shaping procedure described in the in the first paragraph of Sec.~\ref{Sec_Shapes}, using a $7 \times 7$ puckered lattice. Floppy modes were eliminated with extra springs (A) The desired shape (cylinder with subtending angle of $3\pi/2$); (B) relaxed springs in a puckered lattice with three rest lengths ($a_m=1.05a, a_l=1.15a$); (C) lattice with two rest lengths ($a_m=a_l=1.1a$); and (D) lattice with all springs of equal rest length ($a_m=a_l=a$).}
\label{fig_shapes_puckered}
\end{figure*}

Fig.~\ref{fig_shapes_random} presents a few examples for the random lattice with either bending energy or with extra springs at the edges. 
Fig.~\ref{fig_shapes_puckered} is the result of cylindrically shaping the puckered lattice with extra springs. As a comparison we also plot the result of shaping a sheet that has only two different spring rest-lengths, and a sheet with all springs of equal length. Notice how the last two cases completely flatten out, losing their memory of the goal shape. 

All the goal surfaces have zero curvature along the $x$ direction (they thus do not possess Gaussian curvature). To characterize the deviation from the prescribed shape we average all nodes along the $x$ direction, and look at the resulting curve in two dimensions. We then measure $Z_i$, the distance from node $i$ to the corresponding point on the initial surface, allowing rigid body translations and rotations such that the total sum $\sum_i Z_i$ is minimal. We define the error, $\eta$, by 
\begin{equation}
\eta=\sum_i Z_i^2/\sum_i Z_{i0}^2, 
\label{fit}
\end{equation}
where $Z_{i0}$, is the same measure but using the distance between the initial relaxed ``flat" sheet that had zero global curvature and the goal surface. 

Table~\ref{Table_Fit} presents error values for different lattice sizes. For a given lattice size the  more curvature the desired shape has, the worse the fit is (data not shown).
One might expect that in order to get a better fit, all that is needed is to take a larger lattice, but the fit is, in fact, worse (see discussion section for more detail). There is a competition between the number of metastable states available, and the energetic barriers between them. For a small system there are not enough metastable states to imitate the desired shape; for a large lattice, there are many metastable states but the energetic barrier between them is so small that they are not stable. Table~\ref{fit} indicates that between the measured sheets, for a random lattice the preferred lattice is $8\times8$ and for the puckered lattice it is $7\times 7$. 
\begin{table}[tbh]
    \begin{tabular}{ | c | c | c | c | c | c |}
    \hline 
   \multicolumn{6}{|c|}{Random Lattice} \\ \hline
    Lattice size & 4 & 6 & 8 & 10 & 12 \\ \hline
   $\eta$ & 0.56 & 0.36 & 0.10 & 0.13 & 0.30 \\ \hline
\end{tabular}
    \begin{tabular}{ | c | c | c | c | c |}
	\hline
   \multicolumn{5}{|c|}{Puckered Lattice} \\ \hline
  Lattice size & 4 & 7 & 10 & 13 \\ \hline
  $\eta$   & 1.00 & 0.08 & 0.45 & 0.30 \\
    \hline
\end{tabular}
\caption{Effect of lattice size on the average error measure, $\eta$, in fitting to a cylinder of subtending angle of $3\pi/2$ for a random lattices with extra springs at the edges (average over four realizations), and for a puckered lattice with extra springs. The lattices are made as in Fig.~\ref{fig_shapes_random} and Fig.~\ref{fig_shapes_puckered}. }
 \label{Table_Fit}
\end{table}

\begin{figure}[tbh]
\vspace{0.9cm} 
\includegraphics[height=1.5in]{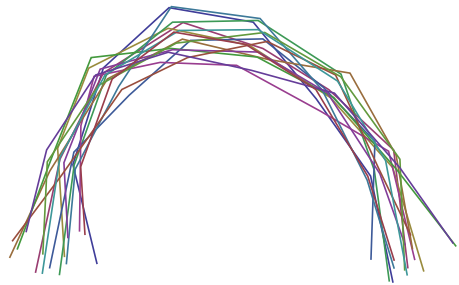} 
\caption{Fifteen random sheets of size $8\times 8$ shaped as half a cylinder. The result presented here is  averaged over the $x$ direction. The error measured by Eq.\ref{fit} gives $\eta=0.06 \pm 0.002$.}
\label{fig_variance}
\end{figure}

To check how reproducible is the result in the random case we took fifteen different random sheets of size $8\times 8$ and shaped them as half a cylinder. The average error value is $\langle \eta\rangle = 0.065$ with a variance of 0.002. The results averaged over the $x$ axis are presented in Fig.~\ref{fig_variance}.

\section{Hysteresis and effective global properties}
\label{sec_Hysteresis}
\begin{figure}[tbh]
\vspace{0.9cm} 
\begin{center}
\centerline{\resizebox{0.4\textwidth}{!}{\includegraphics{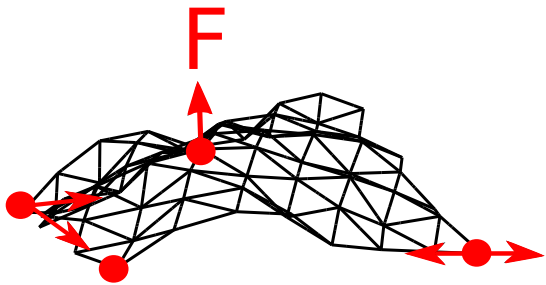}}}
\end{center}
\caption{Representation of the sheet with the center of mass fixed and a force applied on one of the middle nodes--- left front node is completely  fixed, right front node is restricted to a line, left back node is restricted to a plane, and the remaining one is completely free. The force applied on the middle point is causing curvature to the sheet.}
\label{fig_HystCartoon}
\end{figure}

\begin{figure*}[tbh]
\vspace{0.9cm} 
\begin{center}
\centerline{\resizebox{0.8\textwidth}{!}{\includegraphics{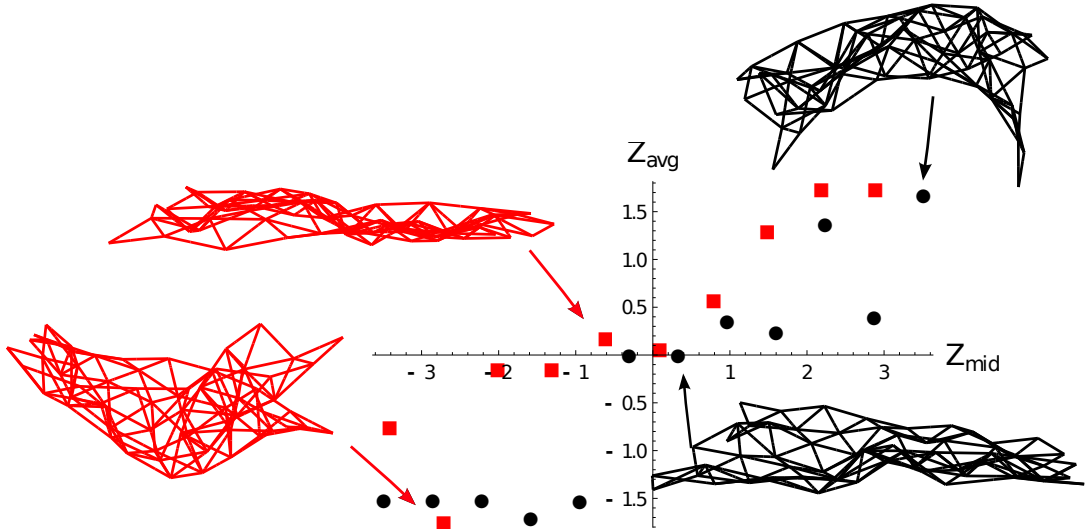}}}
\end{center}
\caption{Hysteresis in a random array of springs of size $8\times 8$ with stiff edges. Forcing the mid point up does not always result in a new surface.  There is an energetic barrier to transform to a new configuration. Plot shows average values and some examples of the relaxed sheet at various points along the loop.}
\label{fig_Hysteresis}
\end{figure*}

Like plastic materials, our system exhibits hysteresis --- the current shape depends on the history of the applied forces. In magnetic materials hysteresis is demonstrated by changing the outer magnetic field in a cyclic fashion and tracking the resulting magnetization. In a similar fashion we changed the applied force cyclically and looked at the average height of the resulting sheet, if there was no memory, the increasing and decreasing forces would trace out the same line.  Since there is memory, we get a loop. We demonstrate hysteresis by the following procedure: (a) We force the midpoint, $Z_{\rm mid}$, upward while pinning three nodes at the edges to define a horizontal plane (pinning one node completely, forcing one to be in a plane, and the third to be on a line). This results in a curved surface. See Fig.~\ref{fig_HystCartoon} for clarification.  
(b) We then release the forced midpoint and minimize the energy. Next, we measure the average height of all nodes in the lattice, $Z_{\rm avg}$. 
(c) Next, we again take the mid point from its current position and force it upwards. We repeat steps b and c until $Z_{\rm mid}$ reaches a few lattice spacings. (The resulting remembered shapes clearly do have Gaussian curvature, unlike the target cylinders of Figs.\ref{fig_shapes_random}A and \ref{fig_shapes_puckered}A). 
(d) Now we force the mid point {\it downward} and repeat up to a few lattice spacings.
(e) We then repeat points (a)-(d) four times. After the second round, variations were small.
For a puckered lattice with stiff edges we get the loop in Fig.~\ref{fig_Hysteresis}. Notice that there are plateaus in several locations. These imply that for a small  increment of force there is no change in the resulting shape, \ie the sheet resists forcing. Each plateau is followed by a jump to a new value. The jump indicates the crossing of the energetic barrier, resulting in a new configuration. 

\begin{figure}[tbh]
\vspace{0.9cm} 
\includegraphics[width=2.5in]{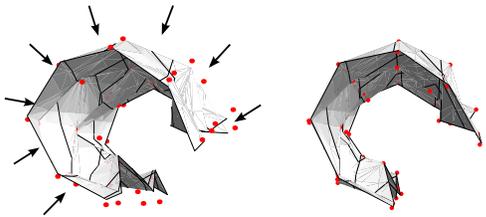} 
\caption{Effect of global forcing on a shaped sheet. A puckered 7x7 sheet with stiff edges was shaped as an almost closed cylinder (represented by red dots in both figures). We applied an outer potential forcing it more inwards (left figure), resulting in six percent deformation. After removing the force, the sheet returns to the original shape (right figure).}
\label{fig_Forced}
\end{figure}

In order to test the robustness of the remembered shapes of Fig.~\ref{fig_shapes_puckered}, we applied an external potential forcing it to curve even more inwards. Up to fifteen percent deformation
 the sheet will go back to the original curved configuration after relaxing the force as can be seen in Fig.~\ref{fig_Forced} (the strain between the initial and final configuration is only 0.008\%). 

In order to have a crosscheck on the numerics we computed the basic properties of one of our networks. The effective global properties of the sheet can be predicted semi-analytically. The lateral bulk modulus $K$ gives the change of lateral pressure $\Delta P$ required to produce a given small change $\Delta A$ in the area of the lattice:
\begin{equation}
K=  -A \frac{d P}{d A} 
\label{bulk_modulus}
\end{equation}
We calculate $K$ numerically by taking a simpler version of the puckered lattice, a lattice that has just one hexagonal pyramid as a repeating unit. (This sheet is less frustrated and therefore somewhat less shapeable than the lattice defined in Sec.~\ref{sec_Model} and used in Figs.~\ref{fig_Model}B and~\ref{fig_shapes_puckered}. It has the advantage of having just nine degrees of freedom and not eighteen). We then uniformly stretch all edges by a small amount (strain of up to 0.3\%) using periodic boundary conditions. By measuring the gain in spring energy $E_s$ under this stretching, we find $K = A (d^2E_s/dA^2) = 0.26$ (where the spring lengths are as in Fig.~\ref{fig_shapes_puckered}, and $k=1$). 

\begin{figure}[tbh]
\vspace{0.9cm} 
\includegraphics[width=2.5in]{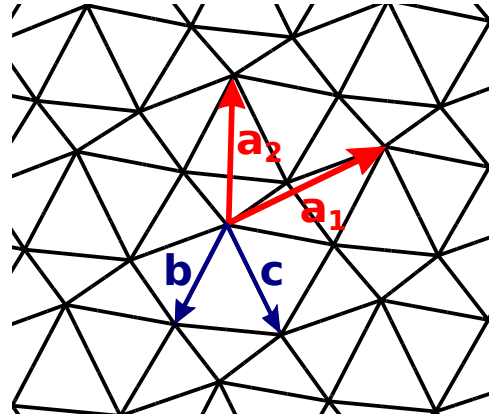} 
\caption{Perspective view of the simplified lattice used for the calculation of bulk modulus in Sec.~\ref{sec_Hysteresis}. Lattice vectors ${\bf a_1}=(l_1,l_2,0),{\bf a_2}=(l_3,0,0),{\bf b}=l_4~{\bf a_1}+ l_5~{\bf a_2} + (0,0,l_6), {\bf c}=l_7~{\bf a_1}+ l_8~{\bf a_2} + (0,0,l_9)$.}
\label{fig_1unit}
\end{figure}

To find an analytic expression in the infinite lattice we proceed as follows --- first we find the ground state of the system. Each unit~cell is completely defined with nine degrees of freedom $l_i$, (three nodes in each unit cell, each of which has three translations), associated with these are nine springs constraining the cell. Since the springs are all relaxed in the ground state, we can find the position of the nodes by solving the nine equations for the springs. The result is shown in Fig.~\ref{fig_1unit} and is similar to the one obtained by numerically minimizing the energy of the periodic sheet. We may then express the energy cost $E_s$ of small deformations of these co-ordinates $\Delta l_i$ in the form $E_s \simeq 1/2~\Delta l_i ~{\bf \hat M}_{ij}~ \Delta l_j $, where the matrix, ${\bf \hat M}$ is given by ${\bf \hat M}\equiv \partial^2 E/\partial l_i \partial l_j |_{l_0}$. 
We then compute ${\bf \hat M}$ around that ground state and express the energy due to spring stretching in terms of $E_s$. 

Applying a small amount of pressure $P$ requires a work $\Delta E_P = P \Delta A$, where $\Delta A$ is the change in area, expressible in terms of $\Delta {\bf l}$. The change of shape induced by $P$ also changes the spring energy $E_s$, also expressible in terms of $\Delta {\bf l}$. Defining  the primitive vectors ${\bf a_1}$ and ${\bf a_2}$ of the unit cell as shown in Fig 9, the area of the cell is evidently $|{\bf a_1 \times a_2 }|$. Thus, the work done by the pressure is given by,
$\Delta E_P=  P ~\Delta|{\bf a_1}\times {\bf a_2}|=  P~ \Delta(l_1 l_3)$. Minimizing the total energy, due to spring stretching and the work done by the pressure, $\partial_{l_j} E_{\rm tot} =\partial_i (E_s + E_P) = 0$ we find, 
\begin{equation}
{\bf l}_j = {\bf l^0}_j -[{\bf \hat M}^{-1}]_{ji} (\partial_i E_P|_{\bf l=l^0}), 
\label{DOF}
\end{equation}
where, ${\bf l}^0$ are the values of the unperturbed lattice.
Given the periodic structure, ${\bf \hat M}$ and ${\bf l_0}$ can readily be found numerically. We now use Eq.~\ref{DOF} to find the area of a unit cell and its derivative with respect to $P$. From this using Eq.~\ref{bulk_modulus}, we can calculate the compressibility, $1/K = -\frac{1}{A} ~dA/dP$ to obtain
\begin{equation}
K = - 1/\left(\frac{1}{l_1}\frac{\partial l_1}{\partial P}+\frac{1}{l_3}\frac{\partial l_3}{\partial P}\right).
\label{compressibility}
\end{equation}
For $a=1, a_m = 1.05 a$, $a_l = 1.15 a$ and $k=1$, we find $K=0.26$ in agreement with the numerical energy minimization calculation. 

For an equilateral lattice the bulk modulus can be found exactly to be $\sqrt{3}/2\,k$ which fits both the semi-analytic calculation and the energy minimization one. The bulk modulus of the puckered lattice is lower than the equilateral one because the middle springs are only slightly strained when a small amount of pressure is applied. The main effect is that the height of the pyramid decreases. Similarly, the bulk modulus of a symmetric puckered lattice (upper drawing of Fig. \ref{fig_Model}B) could also be calculated analytically. In this case, the middle springs of each hexagon play no role at all for the bulk modulus. It is thus similar to the bulk modulus of a honeycomb lattice, which is just one third the bulk modulus of an equilateral lattice \ie $\sqrt{3}/6\,k$. This, again, fits both the semi-analytic calculation and the energy minimization one. As a side comment, notice that this value is slightly higher than the bulk modulus we obtain for the non-symmetric puckered lattice. The reason is that the basis of the hexagonal pyramids in the non-symmetric puckered lattice are slightly out of the plane.

\section{Bistability}
\label{Sec_Bistability}

Where does the shape memory come from? In the examples noted in the introduction, one source of shape memory is simple bistability: the system has two macroscopically different states that are local energy minima.  By exerting macroscopic forces on the system one can cause the configuration to flip to the other minimum. Our lattices also contain such bistable states, which are thus a potential source of the shapeability we seek.  In this section we characterize the bistable states accessed by displacing single nodes such as the pyramid apex of the previous section.  We find that bistability of a node is associated with a geometric feature called angular deficit.  We then investigate the role of these states in the observed shape memory of our sheets. 

Any bistable node has two stable configurations with opposite local mean curvature. These may in principle induce global curvature in the sheet. 
What nodes are bistable? 
There is a correspondence between nodes of positive angular deficit and bistability. Looking at a node and summing the angles around it, the angular deficit is defined as the deviation of that sum from $2\pi$. It is a discrete analogue of Gaussian curvature. The angles at the apex of a hexagonal pyramid sum to less than $2\pi$. Thus, this node has positive angular deficit. The angles at a saddle point sum to more than $2\pi$ and therefore such a node has negative angular deficit. So positive angular deficit corresponds to a node which is a local extremum, if it is a maximum it potentially could be flipped to be a local minimum and vice versa. 

For a lattice of triangles such as ours the angular deficits are subject to a global constraint.  The sum of angular deficits for all nodes of triangular network is unchanged when the nodes are displaced (since the sum of angles over nodes is the same as the sum of angles over their triangles).  This means that changing the angular deficit at one node must change the deficits elsewhere in the network to compensate.

Are all the nodes of positive angular deficit bistable? No. We checked each node for bistability by the following procedure, explained more fully in Sec. \ref{Sec_numerics}. Starting from a given stable state, we flip each node as follows. We determine the plane that best corresponds to the positions of the neighbors. Then we displace the node to its mirror image configuration relative to that plane. We call this the initial trial state.  We then search for a nearby stable state distinct from the starting state. This search proceeds in two steps. 
We first fix all the nodes except the one examined, and determine a nearby energy extremum.  If the node remains separated from its unflipped starting position, we then proceed to vary all node positions and determine a fully stable configuration.   If this stable configuration still remains distinct from the starting unflipped state, we deem this node to be bistable. If on the other hand, the relaxed state reverts to the initial state, we seek other nearby positions of the node that might converge to distinct states.  We return to the initial trial state defined above and displace it by a random amount up to $0.3 a$.  We then test this displaced state for stability as we did for the initial trial state. If the displaced converges to a distinct state, the node is deemed bistable.  If not, we perform another random displacement and test it as before.  If no bistable state is found after 30 such trials, we deem the examined node to be monostable. 

This procedure is adequate for surveying bistable states, but it is not exhaustive.  Since our algorithm to find minima proceeds in discrete jumps, it can fail to find the local minimum corresponding to a given initial state.  Further, this method probes only configurations that can be driven to another stable configuration by displacing a single node. It need not probe all transitions from a given stable state to an adjacent one.

By this procedure we find that in the puckered lattice in the globally flat state all nodes of positive angular deficit (middle of the hexagonal pyramids) are bistable. Most of them stay bistable when cylindrically shaped but not all. In the random lattice there was usually a correspondence between angular deficit and bistability but not always. 

\begin{figure*}[tbh]
\vspace{0.9cm} 
\begin{center}$
\begin{array}{ccc}
\includegraphics[width=5in]{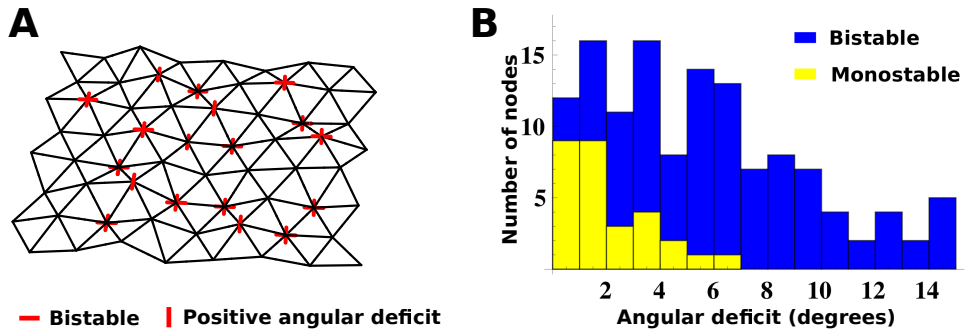} 
\end{array}$
\end{center}
\caption{Bistability in a random lattice with stiff edges  of size 8x8. (A) Bistable  nodes (horizontal lines) and nodes of positive angular deficit (vertical lines) shown on top of the actual nodes. (B) Histogram for the distribution of angular deficit for 252 nodes (in seven different realizations of globally flat random sheets of size 8x8: blue---positive angular deficit and bistable, yellow---positive angular deficit and monostable. Only when the angular deficit is close to zero we get a behavior that deviates from expectation. For large angular deficit, the proportion of monostable nodes falls to zero. We do not present nodes of negative angular deficit since, as mentioned in the text, those were almost always monostable.
}
\label{fig_bistable}
\end{figure*}

These findings imply that bistability is determined partly by the sign of the angular deficit but also by its magnitude and  by the position of the neighbors. Fig.~\ref{fig_bistable} is a result for one random lattice with stiff edges of size $8\times 8$. A horizontal bar at a node indicates positive angular deficit; a vertical bar indicates bistability.  The histogram in Fig.~\ref{fig_bistable} is a distribution of angular deficit for 252 nodes, bistable nodes are dark colored and monostable light colored. One can see that most nodes of positive angular deficit are bistable.  In cases where it is not so, the angular deficit is close to zero. Nodes of negative angular deficit were almost always monostable (out of 252 cases just one instance of negative angular deficit turned out bistable, and the deficit in this case was very close to zero). 

\begin{figure}[tbh]
\vspace{0.9cm} 
\includegraphics[width=2.5in]{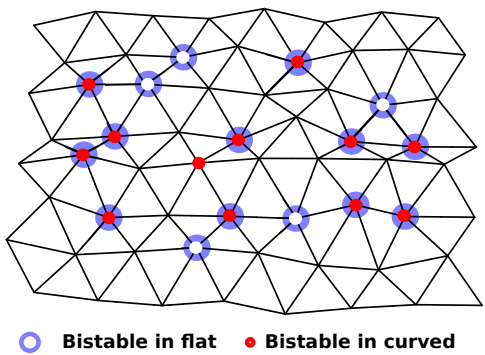} 
\caption{Shaping the sheet changes some of the bistable nodes. An example for  an $8\times 8$ random lattice with stiff edges. Bistable nodes are marked with a blue circle for the flat sheet and with a red dot for the cylindrically curved one.}
\label{fig_bistable_change}
\end{figure}

Shaping a sheet changes some of the nodes from bistable to monostable and vice versa. Fig.~\ref{fig_bistable_change} shows an example of a sheet that started flat with bistable nodes marked by a light circle. We then shaped it cylindrically, as in Fig.~\ref{fig_shapes_random}. The resulting bistable nodes for the cylinder are marked by a dark dot. 

\begin{figure*}[tbh]
\vspace{0.9cm} 
\begin{center}$
\begin{array}{ccc}
\centerline{\includegraphics[width=5in]{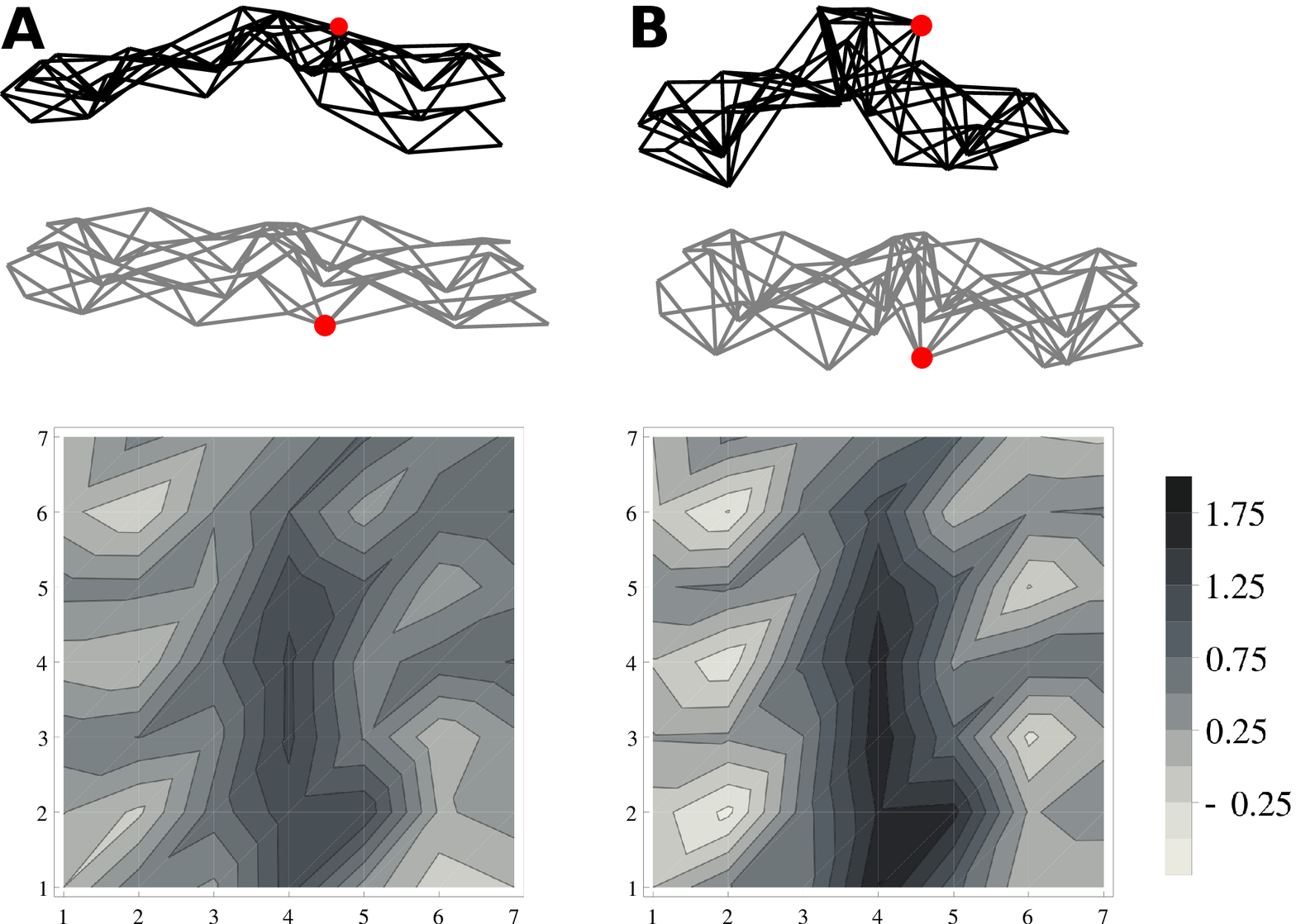}} 
\end{array}$
\end{center}
\caption{Puckered lattice with all bistable nodes pointing down (in gray), and the same lattice with one bistable node (marked by a dot) flipped up (in black). A: sheet with spring lengths as in Fig. \ref{fig_Model}C, B: sheet with longer springs and bigger mismatch ($a_l=1.4a$ and $a_m=1.2a$), resulting in more curvature. Bottom figure presents density plots for the height along the sheet after flipping. The perturbed region extends along a line away from the flipped node.}
\label{fig_Flipped}
\end{figure*}

Let us see how bistability influences the global shape. We take the puckered lattice with extra springs at the edges and flip one of the bistable nodes. Fig.~\ref{fig_Flipped} shows the original flat sheet (light gray) and how it is curved after one node is flipped (black). Taking longer springs in the middle of the hexagons results in larger curvature (right figure). 

\begin{figure}[tbh]
\vspace{0.9cm} 
\centerline{\includegraphics[width=3.5in]{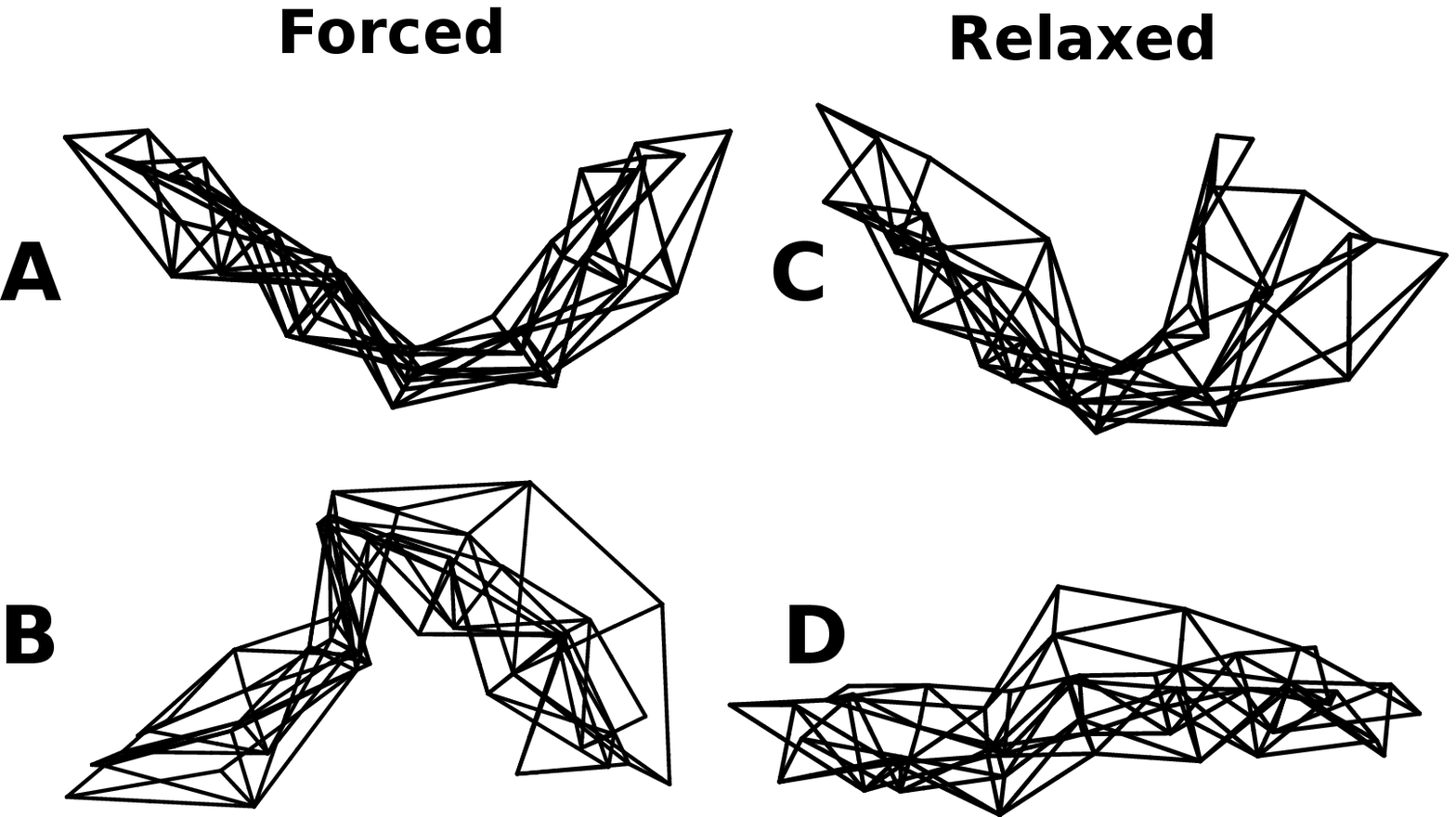}\hspace{0.5cm} }
\caption{Puckered lattice used in Fig.~\ref{fig_Flipped}A. It is either forced to curve up (A) or down (B) as described in the text. If the force is removed, the shape retains some curvature when bistable nodes point outward on the curved surface (C), but not when they point inwards (D).}
\label{fig_UpDown}
\end{figure}

The local curvature of the bistable nodes dictates possible global curvatures for the entire sheet. Fig.~\ref{fig_UpDown}A demonstrates how a flat puckered  sheet with all hexagons pointing down (all having local mean downward curvature) can be forced to curve such that it has global mean downward curvature. After removing the force, it will stay curved. On the other hand, Fig.~\ref{fig_UpDown}B indicates that forcing it in the opposite direction, \ie trying to impose upward curvature, does not work --- the sheet flattens once the force is removed. 

Bistability is important, but it is not the only factor that determines the shape. There are multiple stable shapes for the same configuration of bistable nodes, as shown in Fig.~\ref{fig_UpDown}. Here an initially flat state was forced to bend by constraining the middle line and forcing the two edges up. The resulting configuration (A) did not undergo
any flips in the bistable nodes, nor did any nodes flip when the force was removed (C). The set of bistable nodes remained unchanged in both (A) and (C). Further, (C) was robust to
perturbations. It returned to the configuration shown when fifteen percent random displacements in the node positions were imposed. 

\begin{figure}[tbh]
\vspace{0.9cm} 
\centerline{\includegraphics[width=3.5in]{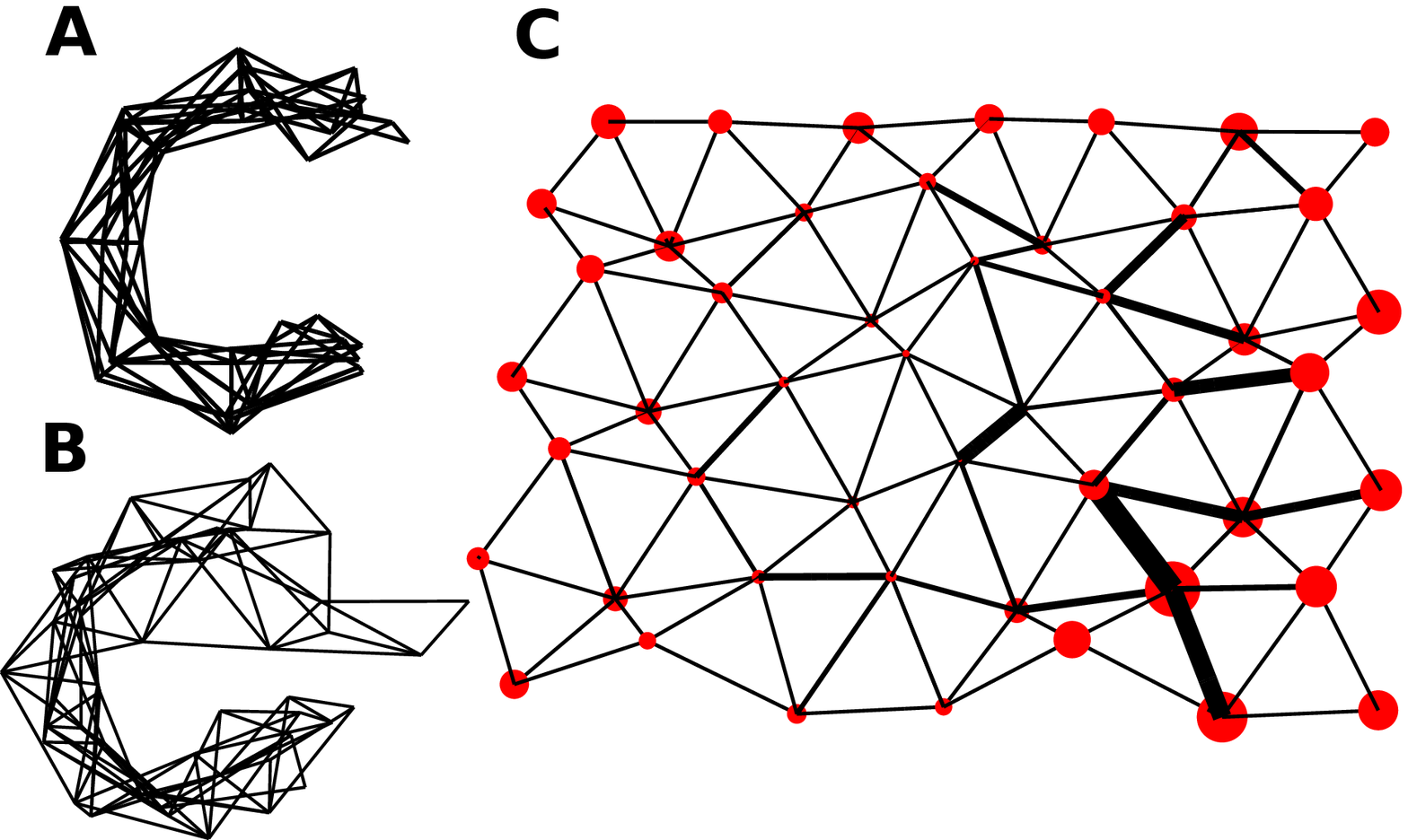}\hspace{0.5cm} }
\caption{(A) Puckered lattice used in Fig.~\ref{fig_Forced}. It is then forced beyond its elastic limit and, when relaxed, finds a new configuration (B). The normalized change in dihedral angles between each pair of triangles (controlling bending energy) is represented by the width of the lines in (C), the displacement of a node is given by the size of the dots .}
\label{fig_IPR}
\end{figure}

To characterize the non-locality, we look at a cylindrically shaped puckered sheet and force it even more inwards, same as was done in Fig.~\ref{fig_Forced}. This time we push it just above the limit of elasticity, so it does not recover. We would like to define the change between this state and the previous one. Is the change very local? Did just one node flip? Or did all of them move? A good way to measure locality is to look at the Inverse Participation Ratio (IPR), defined by: 
\begin{equation}
{\rm IPR} = \frac{1}{\sum_i \psi_i^4},
\label{IPR}
\end{equation}
where $\psi_i$ is the change in dihedral angle between every pair of adjacent triangles, normalized such that $\sum_i \psi_i^2 = 1$. The IPR gives 1 if the change is localized in one spring, and $N$ if it is spread equally over all springs. In the above deformation for a $7\times 7$ lattice we get  IPR=12. \ie about 12 sites accounted for most of the displacement after the removal of the force. None of the bistable nodes have flipped, and checking to see where the largest change occurred, we find that it happened at a saddle point. Fig.~\ref{fig_IPR} presents the initial cylinder and the relaxed one after forcing beyond elasticity in (A) and (B) respectively.  Fig.~\ref{fig_IPR}C shows the amount of displacement (dot size) of each node and the change in dihedral angles (line width) between A and B displayed on the ``flat" initial sheet. Observe that the biggest change in angles is not around just one node, but  also not spread on the entire shape, but rather localized around a few nodes. 

\section{Numerics}
\label{Sec_numerics}

\begin{figure}[tbh]
\vspace{0.9cm} 
\includegraphics[width=2.5in]{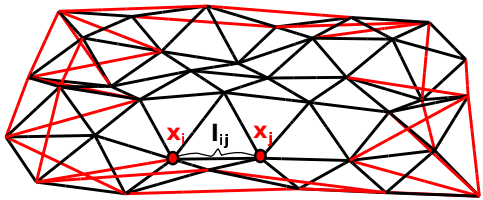} 
\caption{An example of a random triangular lattice with nodes $x_\alpha$ and $x_\beta$ connected by springs of rest length $l_{\alpha\beta}$. Red lines represent extra springs at the edges.}
\label{fig_Numerics}
\end{figure}

The lattice definition above specifies the spring basic energy $E_0({\bf x}_1, ... {\bf x}_n)$ as a function of node positions, $E_0=\frac12 k \sum_{[\alpha,\beta]} (|{\bf x}_\alpha-{\bf x}_\beta|-l_{\alpha\beta})^2$, where ${\bf x}_\alpha$ is the position of the  $\alpha$ node, and the sum goes over all springs (see Fig.~\ref{fig_Numerics}). The rest length, $l_{\alpha\beta}$, depends on the model, as explained in Sec.\ref{sec_Model}. Given a numerical formula for $E_0$ we must determine the node positions that minimize this energy. Standard numerical methods give an iterative prescriptions for approaching this minimum, as discussed below. 

To eliminate floppy modes we add bending energy or extra springs. Extra springs are added to nodes at the edges that have less than six neighbors. We connect them to their next to nearest neighbors using a spring constant that is smaller by a factor of 10, \ie $k/10$. The rest length of these springs is chosen such that they are relaxed in the original configuration. Bending energy is given by $E_b=C~ \sum_{[\mu, \nu]}1/(1.1+\hat{\bf n}_\mu \cdot \hat{\bf n}_\nu)$, where $\hat{\bf n}_\mu$ is the normal to the surface of the $\mu$th triangle, and the sum goes over neighboring triangles. This particular form guarantees that the energy increases sharply as the angle between triangles gets closer to $\pi$. This discourages triangles from simply folding onto their neighbors. The constant $C$ was chosen to be $C =k a^2/3000 $ such that~$C \ll k a^2$.

If our sheets were physical objects in the real world they would find the closest minimum to the initial configurations. 
In this qualitative study we used the standard, nonlocal minimization methods, since these were faster and captured the qualitative features. Specifically, we used the FindMinimum command in Mathematica \cite{Mathematica}. We tested a few methods under FindMinumum --- conjugate gradient, Newton and quasi-Newton. The results didn't differ qualitatively, producing the same average error values. However, convergence times were  longer than the general procedure. We thus used the FindMinimum without specifying any method.  These minimization schemes do not necessarily scan the energy landscape in a continuous fashion. 
The ``springback test" of Fig.~\ref{fig_Forced} above confirms that the shaped configuration is a robust minimum. 
The numerical calculation of the bulk modulus provided additional validation of the numerics and the input energy formulas used in the numerics.

 For the calculation of bistable nodes we went over each node in the lattice, each time fixing all nodes but one. We used the function FindRoot in {\it Mathematica}\cite{Mathematica} which implements Newton's method to find the root of a set of equations. In this instance we used it to find an extremum. It requires an initial guess for which we take the mirror image of the free node plus a small random number taken from the interval $[-0.3,0.3]a$ . The mirror plane was calculated by finding a plane which is the closest to the six neighbors of the free node. If the extremum point is in the vicinity of the original node (within $\pm 0.05a$) we say it is the same position and go on to look at another initial guess. We do this for up to 30 times. If in all of those tries we didn't find a second stable configuration we conclude that the point is monostable. 
We then take the list of nodes that are suspected as bistable and for each one relax the sheet globally using {\it Mathematica's} FindMinimum. If the position of the flipped node is different from the original one by more than $\pm0.05 a$ we say that it is truly bistable. 

\section{Discussion}
\label{sec_Discussion}

This study was based on the notion that the shape memory seen in crumpled paper is distinctive and robust because of its two-dimensional connectivity.  We aimed to capture this form of shape memory by a minimal system embodying this two-dimensional connectivity along with the local bistability of a crumpled sheet, using a simple lattice of springs.  Remarkably,  this lattice showed significant shape memory in empirical numerical studies.  Indeed, the resulting shapes resembled shapes seen when one physically shapes crumpled paper.  The remembered shapes were robust: even when they were deformed significantly by external forces, they returned to their remembered shape when these forces were removed.  In this section we examine the origin and potential significance of this intriguing behavior.  We focus on the puckered lattice configuration, since it is the simplest system that shows the shapeability.  

\subsection*{Plastic vs recoverable shapeability}  

Macroscopically, the shapeability of our sheet is no different from that of a malleable piece of metal, such as a coat-hanger wire.  When either of these materials is forced into a given shape, it retains that shape.  If it is forced moderately from the retained shape, it deforms elastically, returning to that shape when the force is released.  In this sense it is as shapeable as our system.  The distinctive aspect of our sheet lies in the nature of the microscopic changes that allow retention of a shape.  In the malleable metal the new shape arises because of plastic deformation.  Planes of atoms making up the metal crystal slide past each other, to reach another stable state in which there has been a net relative motion of the atoms.  In making a macroscopic deformation, this process is repeated so that the material displacement between two given atoms may grow to indefinite size.  The microscopic variables that describe plastic deformation must thus cover an indefinite range. Such deformations are not recoverable.  That is, the original arrangement of the atoms cannot be recovered by the type of external forcing that led to plastic deformation. In our sheet, by contrast, the microscopic variables may be taken to be the spring lengths.  All the retained shapes of the sheet are defined by limited changes of these lengths of the order of a fraction of a lattice length.  Because of this, the deformations are recoverable.  One can return to the initial microscopic state by applying a suitable force.  For example, one may force the nodes into a plane to create a unique flat reference state.  By contrast, one cannot restore a bent wire to its initial straight shape with all the atoms in their initial positions.

\subsection*{Shapeability and local bistability}  

Any locally stable configuration of a mechanical system implies a local minimum of its potential energy.  Since our system has multiple stable states ---\eg flat {\it vs} curved--- it must have multiple local energy minima.  Moreover, these minima are coupled to macroscopic curvature and are selectable by imposing macroscopic curvature profiles.
				
Our system was designed to have many local energy minima.  First, it is constructed to be hyperstatic, so that there are no free motions degenerate in energy.  Second, it was constructed to have an extensive set of bistable states associated with individual nodes of the lattice.  We found empirically that simply having such bistable states was not sufficient for shape memory.  Instead, it was necessary that the state of one bistable node affect the stable positions of the other bistable nodes. Thus deforming the network causes bistable nodes to become stable and vice versa (Fig. \ref{fig_bistable_change}). Likewise, a given external force may collectively destabilize a family of bistable states due to their interaction. 
				
We expect any two-dimensional sheet to have such cooperativity. In a smooth, unstretchable sheet, the Gaussian curvature must vanish everywhere: one principal curvature must vanish at every point, and the two uncurved directions extending from any point must form a straight line to the boundary \cite{Parker1977}.
Real sheets differ from this ideal case.  They can stretch and fold, thus weakening these constraints.  Still, the requirement of remaining as a continuous sheet imposes strong constraints on the energy landscape. Thus the minima of interest in our sheet are expected to be co-operative, involving multiple nodes.  For example, the remembered states of cylindrical curvature observed in our study involve such cooperativity (Fig. \ref{fig_IPR}).  The curvature at a given point is shared by several nodes.
				
The above picture leads us to expect a strong connection between the deformation into a remembered shape and an associated flipping pattern of the bistable states.  We did observe some relationship between the direction of imposed curvature and the flipping of bistable states, as described in Fig.~\ref{fig_Flipped}.   However, the relationship was far too weak to explain the robust retention of shapes that we observed.  Our system was able to retain strongly curved states without {\it any} change of the bistable states we monitored.  
				  
By deforming a shaped sheet beyond the threshold of irreversibility, we got some indication of the nature of the minima.  When the imposed deformation force was pushed just past the threshold, we observed a small discontinuous displacement.  Some of this displacement remains after the force is removed.  This displacement has moved the system from one energy minimum to another nearby minimum.  The shift was accomplished with no flipping of our bistable nodes, as noted above. Instead, the shift was a pattern of spring deformations concentrated along one row of nodes.  This motion confirms our expectation that the energy minima responsible for shape memory are not local but co-operatively stored by multiple nodes and springs.  Characterizing these moves further will be important for understanding these shapeable lattices. 
				
\subsection*{Scalability}
Our study gives information about how the shape memory depends on the number of nodes in the lattice.  When we simply created a larger lattice with the same local structure, the shape memory decreased.  A larger sheet bent through a given angle relaxes nearly completely while a smaller sheet remains bent.  This behavior is natural in the continuum limit.  Any mechanical sheet when bent with a curvature sufficiently smaller than its inverse thickness, must respond elastically, and thus reversibly. Conversely,  shaping behavior of a sheet on the scale $L$ requires a non-elastic, irreversible response for curvatures of order $1/L$. This suggests that the effective thickness should be of order $L$ to retain shapeability.  The 8 x 8 sheets of our main study satisfied this criterion. They had a root-mean-square thickness of roughly five percent of their width. To expand the thickness in proportion to $L$ as $L\rightarrow \infty$ cannot be achieved with simple lattices like those studied here.  Instead, one would need to introduce structure on increasingly large wavelength scales, with the long wavelengths supplying the needed thickness on the largest scales.  We note that crumpled sheets have bendable elements on many length scales \cite{Blair2005} so that their effective thickness grows with their size.  
					
Another potential way to modify the lattice so that it remembers weak curvature is to reduce the distance between the bistable node positions \eg by reducing the height of the pyramids in Fig. \ref{fig_Model}.  Reducing this distance must tend to reduce the amount of deformation (\ie curvature) needed to produce a flip.

In view of these ways to enhance shapeability, our observed reduction in shapeability  with $L$ using our constant lattice geometry does not appear insurmountable.

\subsection*{Compound curvature}
Notably lacking from our study was compound curvature. Our main studies were confined to cylinder-like shapes with curvature in only one direction.  This simple curvature was sufficient to demonstrate shapeability.  Still, such shapes are very limited.  In particular they are far more limited than the general three-dimensional shapes formable using crumpled sheets or origami shapes such as the ``water bomb" \cite{Greenberg2011}.  The hysteretic shapes of Fig. \ref{fig_Hysteresis} showed some compound curvature as well as those of Fig.~\ref{fig_Flipped}. We did not systematically attempt such shapes, for the reason noted above.  Any smooth sheet with compound curvature must undergo large variations in the spatial distance between material points, \ie\ large and inhomogeneous strain.  Crumpled sheets satisfy this constraint by folding.  Folding allows large distances in the material sheet to span only small distances in space.  Our lattices were not amenable to folding; thus, we did not expect them to remember shapes with compound curvature.  However, generalizing our lattices to allow folding should permit the lattices to adopt shapes with substantial compound curvature.

\subsection*{Connection to other material memories}

Shapeability is a form of memory, as emphasized above.  Several other forms of material memory have received wide attention in recent times, in addition to those mentioned in the Introduction.  Examples are the classic spin-glass associative memory of Hopfield \cite{Hopfield1982}, the sheared colloidal dispersions of Pine and Chaikin\cite{Pine2008}, and the selectable crystallization of a ``magic soup" of components of Murugan et al \cite{Arvind2014}.  The question naturally arises how the shapeable sheets studied above are related to other forms of memory.	
							
Any physical system that functions as a memory associates a (large) set of configurations $\{c\}$ with a (small) set of target configurations $\{g\} \subset \{c\}$.  The association means that for each target configuration $g_i$ there exists a set of other configurations $\{c\}_i \subset \{c\}$ such that any initial configuration $C \in \{c\}_i$ evolves into $g_i$ and remains at $g_i$. The number of target configurations $g_i$ can range from one to a large number.  The number of initial configurations $\{c\}_i$ leading to a given $g_i$ may also range widely, from a single configuration---$g_i$ itself--- to a large fraction of the possible configurations.  For example, an array of $N$ decoupled magnetic bits, has a capacity of $2^N$ target states, but the set of initial bit patterns $\{c\}_i$ corresponding to a given target bit pattern $g_i$ consists of only the single configuration $\{c\}_i = g_i$.  Conversely, a single ideal ferromagnet whose atomic spins are forced into a given pattern relaxes to one of only two states: the ``up" and the ``down" ground states.  Here there are only two $g_i$ and virtually all the configurations $c$ belong to either $\{c\}_1$ or $\{c\}_2$.  
				
In several of these systems, \eg the spin glass memory and the magic soup, the memories are pre-determined or {\it instilled} by a separate process.  This instillment does not play a role in the shapeability explored in this work.  The shapeability arises from generic features of the structure; the desired shapes were not explicitly programmed into the lattice.

An ideal shapeable material can assume a wide range of coarse-grained geometric forms.  Thus an ideal shapeable sheet would be able to approximate any smooth profile of compound curvature, such as a U-channel, a bowl or a saddle shape. The process of selecting a target state consists of forcing the sheet into a shape similar to that of the target state.  The memory consists of the retention of this form under perturbations.  The set of deformations that return to the target state are the $\{c\}_i$ for this shape $g_i$.  The material can retain a large range of possible shapes; thus the range of $\{c\}_i$ selecting a given $g_i$ is a small fraction of this total range.  Since any given region may in principle be shaped independently, the number of possible memories is potentially proportional to the number of configurations of the system and exponential in the number of degrees of freedom.  The capacity of the spin-glass memory, by contrast, is simply proportional to the number of degrees of freedom\cite{Hopfield1982}

\subsection*{Physical realizations}
The utility of the sheets studied here depends on physical realizations.  The simulations presented above provide encouragement that networks of real nodes and springs will show shape memory, though these simulations give only a qualitative representation of a real network.  In a real network bending elasticity is needed in order to prevent unconstrained modes of motion, but our simulated bending elasticity was not especially realistic. A wide range of physical implementations would be consistent with the qualitative properties of our simulation.  In particular, the network could be molded or cast as a single piece of plastic or metal.  Our simulations made little attempt to optimize the geometry of the structure.  Thus there is great scope for improved shapeability.

\section*{Conclusion}

Deforming two-dimensional elastic manifolds into three dimensions typically induces a reinforcing network of ridges and vertices \cite{Witten:2007fk}.  In this study we have investigated how this co-operative response might influence a manifold containing local energy minima.  We speculated that the induced network might couple the local energy minima so as to create remembered shapes.    Our exploratory lattice models made to test this mechanism indeed showed a modest but unambiguous shapeability.  
Thus they demonstrate that extensive shapeability is achievable without plastic deformation and without designing the material to create specific shapes. They thus suggest a new strategy for creating deformable, reconfigurable objects.  Further, this mechanism may account for the extensive shapeability seen in everyday crumpled sheets of paper or plastic.

To understand how the shape memories are stored, one must understand the constraints that define a given energy minimum and that dictate the transitions between minima.  We have only begun to explore these minima.  It appears feasible that lattices like those studied here can be developed into a generic form of shapeable material.  Our work towards both of these goals is in progress.  

\begin{acknowledgments}
We are grateful to Jin Wang, Efraim Efrati, Arvind Murugan, Matan Ben-Zion, Sidney Nagel and Martin Van Hecke for fruitful discussions.  Matthew Pinson provided a valuable critique of the manuscript. N. O. was supported by a Kadanoff-Rice fellowship from the University of Chicago's Materials Research Science and Engineering Center, funded by the National Science Foundation under award number DMR-0820054. 

\end{acknowledgments}


\end{document}